\input{aipcheck}

\documentclass[
    ,final            % use final for the camera ready runs
  ]
  {aipproc}

\layoutstyle{8x11single}

\begin{document}

\title{Combined analysis of the decays $\tau^{-}\to K_{S}\pi^{-}\nu_{\tau}$ and $\tau^{-}\to K^{-}\eta\nu_{\tau}$}


\author{Sergi Gonz\`{a}lez-Sol\'{i}s}{
  address={Grup de F\'{\i}sica Te\`orica (Departament de F\'{\i}sica) and
                Institut de F\'{\i}sica d'Altes Energies (IFAE), Universitat
                Aut\`onoma de Barcelona, E-08193 Bellaterra (Barcelona), Catalunya.}         
}

\begin{abstract}

The potential of performing a combined analysis of the strangeness-changing decays $\tau^{-}\to K_{S}\pi^{-}\nu_{\tau}$ and $\tau^{-}\to K^{-}\eta\nu_{\tau}$ for unveiling the $K^{*}(1410)$ resonance pole parameters is illustrated. Our study is carried out within the framework of Chiral Perturbation Theory, including resonances as explicit degrees of freedom. Resummation of final state interactions are considered through a dispersive parameterization of the required form factors. A considerable improvement in the determination of the pole position with mass $M_{K^{*}(1410)}=1304\pm17$ MeV and width $\Gamma_{K^{*}(1410)}=171\pm62$ MeV is obtained.

\end{abstract}

\maketitle

%%%%%%%%%%%%%%%%%%%%%%%%%%%%%%%%%%%%%%%%%%%%
%% MAINMATTER
%%%%%%%%%%%%%%%%%%%%%%%%%%%%%%%%%%%%%%%%%%%%

\section{Introduction}

Hadronic decays of the $\tau$ lepton provide a clean framework for the study of QCD in its non-perturbative regime. Exclusive decays, in which the final state hadrons are known, constitute an ideal scenario for understanding the hadronization of QCD currents as well as for determining the physical parameters, such as the mass and the width, of the intermediate resonances that drive the decays. In this work, we (re)analyze the experimental measurement of the invariant mass distribution of the decay $\tau^{-}\to K_{S}\pi^{-}\nu_{\tau}$ together with the most recent available spectrum of the $K^{-}\eta$ decay mode both released by the Belle Collaboration \cite{Epifanov:2007rf,Inami:2008ar}. The former has been studied in detail in Refs. \cite{Jamin:2008qg,Boito:2008fq,Boito:2010me}, improving the determination of the resonance parameters of both the $K^{*}(892)$ and its first radial excitation $K^{*}(1410)$, while the later, with a threshold above the $K^{*}(892)$ dominance, has been recently tackled in Ref. \cite{Escribano:2013bca} obtaining the $K^{*}(1410)$ properties which appeared to be in accordance with those of the $K_{S}\pi^{-}$ decay channel. In order to deepen our knowledge of the $K^*{}(1410)$ resonance parameters we performed in Ref. \cite{Escribano:2014joa} a combined analysis of both decays, whose summary is the main purpose of the present work.\\
In this talk we address the following topics: in section 2, the construction of the participant vector form factor is discussed in detail while the scalar form factors are borrowed from Ref. \cite{Jamin:2001zq}, and both enter into the corresponding differential decay rate distributions which afterwards will be used to fit the experimental spectra. Our fit results are presented and discussed in section 3 where we emphasize the necessity of measuring the interesting $\tau^{-}\to K^{-}\pi^{0}\nu_{\tau}$ decay channel for investigating possible isospin violations in the low-energy form factor slope parameters. Section 4 is devoted to our conclusions.

\section{Form factors representations}

The theoretical expression for the differential decay rate distribution of the decay $\tau^{-}\to K_{S}\pi^{-}\nu_{\tau}$ is written as

\begin{eqnarray}
\label{spectral function}
\frac{{\rm d}\Gamma(\tau^-\to K_S\pi^-\nu_\tau)}{{\rm d}\sqrt{s}} &\,=\,&
\frac{G_F^2 M_\tau^3}{96\pi^3s} S_{EW}\Big|V_{us}f_+^{K_S\pi^-}(0)\Big|^2
\biggl(1-\frac{s}{M_\tau^2}\biggr)^2q_{K_S\pi^-}(s) \\
\vbox{\vskip 8mm}
&\times& \biggl\{\left(1+\frac{2s}{M_\tau^2}\right)q_{K_S\pi^-}^2(s)\Big|\widetilde{f}_+^{K_S\pi^-}(s)\Big|^2+\frac{3\Delta_{K_S\pi^-}^2}{4s}
\Big|\widetilde{f}_0^{K_S\pi^-}(s)\Big|^2\biggr\} \,, \nonumber
\end{eqnarray}
where
\begin{equation}
\label{definitions}
q_{PQ}(s) \,=\, \frac{\sqrt{s^2-2s\Sigma_{PQ}+\Delta_{PQ}^2}}{2\sqrt{s}}\,,\quad
\Sigma_{PQ} \,=\, m_P^2+m_Q^2 \,, \quad
\Delta_{PQ} \,=\, m_P^2-m_Q^2 \,,
\end{equation}
and
\begin{equation}
\widetilde{f}_{+,0}^{PQ}(s) \,\equiv\, \frac{f_{+,0}^{PQ}(s)}{f_{+,0}^{PQ}(0)}
\end{equation}
are form factors normalised to unity at the origin. The respective formula for the $\tau^{-}\to K^{-}\eta\nu_{\tau}$ can be found in Ref. \cite{Escribano:2013bca}. The main advantage of this parameterization is that the scalar form factor $f_{0}(s)$ corresponds to the $S$-wave projection of the $K\pi$ system whilst the vector form factor $f_{+}(s)$ is the $P$-wave component. Regarding the global normalization, we employ $|V_{us}f_+^{K_S\pi^-}(0)|=0.2163(5)$ \cite{Antonelli:2010yf} and $S_{EW} = 1.0201$ \cite{Erler:2002mv} accounting for the electroweak correction.

The initial setup of our approach to describe the required vector form factor (VFF) is within the context of resonance chiral theory \cite{Ecker:1988te} which after imposing the asymptotic falloff as $1/s$ it reads, for the case of the $K\pi$ system, as

\begin{equation}
f_{+}^{K\pi}(s)=\frac{m_{K^{*}}^{2}+\gamma s}{m^{2}_{K^{*}}-s}-\frac{\gamma s}{m^{2}_{K^{*\prime}}-s},
\label{FF1}
\end{equation}
where $K^{*}=K^{*}(892)$ and $K^{*\prime}=K^{*}(1410)$ are the resonances explicitly considered in our model and $\gamma$ is a dimensionless parameter that weights the relative importance of the second resonance with respect to the first one. Looking at Eq.(\ref{FF1}) one immediately realizes that this description breaks down at $s=m^{2}_{K^{*(\prime)}}$ when the intermediate resonance(s) are on-shell. The most common way to cure this limitation is by taking into account possible rescattering effects of the final state hadrons. These unitarity corrections are incorporated by resumming the whole series of self-energy insertions in the propagator. Finally the reduced form factor takes the form \cite{Boito:2008fq}

\begin{equation}
\label{FF2}
\widetilde{f}_+^{K\pi}(s) \,=\, \frac{m_{K^*}^2 - \kappa_{K^*}\,
\widetilde{H}_{K\pi}(0) + \gamma s}{D(m_{K^*},\gamma_{K^*})} -
\frac{\gamma s}{D(m_{K^{*\prime}},\gamma_{K^{*\prime}})} \,,
\end{equation}
where
\begin{equation}
\label{Dden}
D(m_n,\gamma_n) \,=\, m_n^2 - s - \kappa_n \widetilde{H}_{K\pi}(s) \,,
\end{equation}
and
\begin{equation}
\label{kappa}
\kappa_n \,=\, \frac{192\pi}{\sigma_{K\pi}(m_n^2)^3}\frac{\gamma_n}{m_n} \,.
\end{equation}

The scalar one-loop integral function $\widetilde{H}_{K\pi}(s)$ is defined in Ref.\cite{Jamin:2006tk} and since the $K^{*(\prime)}$ resonances can decay into both $K^{0}\pi^{-}$ as well as $K^{-}\pi^{0}$ channels we have considered appropriate to employ an isospin average form, that is

\begin{equation}
\label{HtKpi}
\widetilde{H}_{K\pi}(s) \,=\, \frac{2}{3}\,\widetilde{H}_{K^0\pi^-}(s) +
                              \frac{1}{3}\,\widetilde{H}_{K^-\pi^0}(s),
\end{equation} 
where $\sigma_{K\pi}(s)$ are the phase space functions given by $\sigma_{K\pi}(s)=\lambda(s,m_{K}^{2},m_{\pi}^{2})/s$ where $\lambda(x,y,z)$ is the so-called K\"{a}llen function. We want to emphasize here that $m_{n}$ and $\gamma_{n}$ are nothing but the unphysical "mass" and "width" parameters to be differentiated from the physical ones which will be determined later from the pole position in the complex plane.\\
Our form factor should satisfy analyticity. A two-meson form factor is an analytic function everywhere in the complex plane except for the branch cut starting at the the two-particle production threshold ($K\pi$ in our case) where an imaginary part is developed. Then, analyticity relates the imaginary and the real part of the form factor through a dispersion relation. We will only consider elastic $K\pi$ rescatterings and then, following the prescriptions of Ref. \cite{Boito:2008fq}, we will employ in this work a three-times subtracted dispersive representation of the form factor,

\begin{equation}
\label{dispersive VFF}
\widetilde{F}_+^{K\pi}(s) \,=\, \exp\Biggl[\, \alpha_1\frac{s}{M_{\pi^-}^2} +
\frac{1}{2}\alpha_2\frac{s^2}{M_{\pi^-}^4} +\frac{s^3}{\pi}
\int\limits_{s_{K\pi}}^{s_{\rm cut}} {\rm d}s'
\frac{\delta_1^{K\pi}(s')}{(s')^3(s'-s-i0)} \,\Biggr].
\end{equation}
The form factor written as in Eq. (\ref{dispersive VFF}) suppresses the less-known high-energy region where the possible inelastic effects, starting at the $K^{*}\pi$ threshold, are already present. The associated error has been estimated and incorporated as systematic by varying the cut-off $s_{cut}$. The two subtraction constants $\alpha_{1,2}$ are related to the slope parameters appearing in the low-energy expansion of Eq. (\ref{FF1})

\begin{equation}
\label{slope parameters}
\widetilde{f}_+^{K\pi}(s) \,=\, 1 + \lambda_+^{'}\frac{s}{M_{\pi^-}^2} +
\frac{1}{2}\lambda_+^{''}\frac{s^2}{M_{\pi^-}^4} + \ldots \,.
\end{equation}
Explicitly we have $\lambda_{+}^{\prime}=\alpha_{1}$ and $\lambda_{+}^{\prime}=\alpha_{2}+\alpha_{1}^{2}$. These parameters will be determined from the fit.\\
Finally, the phase of the form factor appearing in Eq. (\ref{dispersive VFF}) is calculated from the relation 
\begin{equation}
\label{del1Kpi2}
\tan\delta_1^{K\pi}(s) = \frac{{\rm Im}\widetilde{f}_+^{K\pi}(s)}
                              {{\rm Re}\widetilde{f}_+^{K\pi}(s)}.
\end{equation}

All we have discussed above was about the $K\pi$ VFF. The $K^{-}\eta$ VFF is found to be just $\cos\theta$ times the $K^{-}\pi^{0}$ VFF, where $\theta$ is the $\eta-\eta^{\prime}$ mixing angle with a value of $\theta=-(13.3\pm1.0)^{\circ}$ \cite{Ambrosino:2006gk}. 

The required scalar form factors appearing in Eq. (\ref{spectral function}) were worked out in Ref. \cite{Jamin:2001zq} through a careful treatment of the inelasticities appearing at the $K\eta^{(\prime)}$ thresholds and performing a coupled channel analysis. We borrowed their results in this work.

\section{Joint fits to $\tau^{-}\to K_{S}\pi^{-}\nu_{\tau}$ and $\tau^{-}\to K^{-}\eta\nu_{\tau}$ Belle data}

\begin{figure}[thb]
\vspace*{1.25cm}
\includegraphics[scale=0.85]{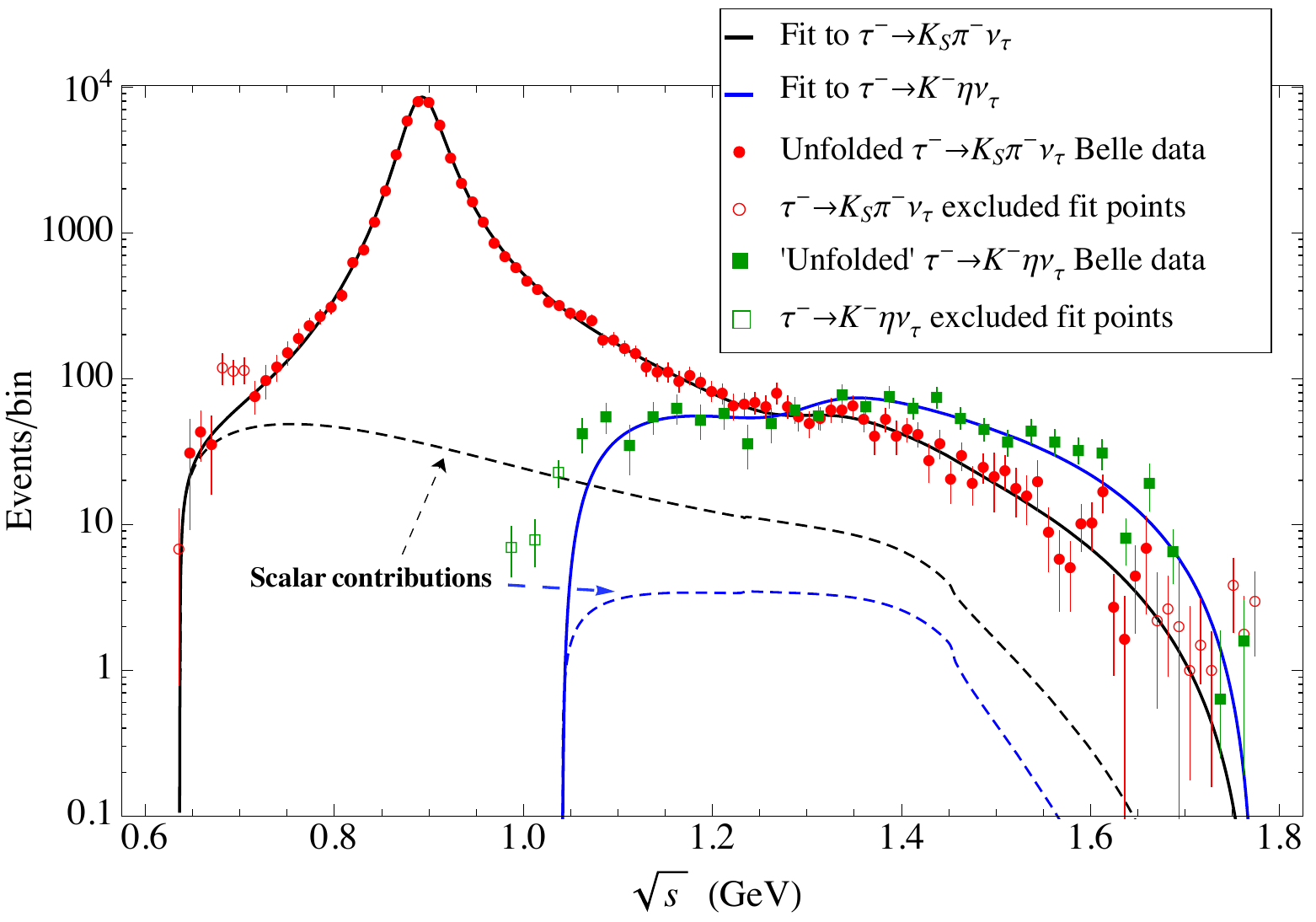}
\caption{\label{fig:Spectrum} \small{Belle $\tau^-\to K_S\pi^-\nu_\tau$
(red solid circles) \cite{Epifanov:2007rf} and $\tau^-\to K^-\eta\nu_\tau$
(green solid squares) \cite{Inami:2008ar} measurements as compared to our best
fit results (solid black and blue lines, respectively) obtained in combined
fits to both data sets, as presented in table.\ref{tab:a}. Empty
circles (squares) correspond to data points which have not been included in
the analysis. The small scalar contributions have been represented by black
and blue dashed lines showing that while the former plays a role for the $K\pi$
spectrum close to threshold, the latter is irrelevant for the $K\eta$
distribution.}}
\end{figure}

\begin{figure}[thb]
\vspace*{1.25cm}
\includegraphics[scale=0.56]{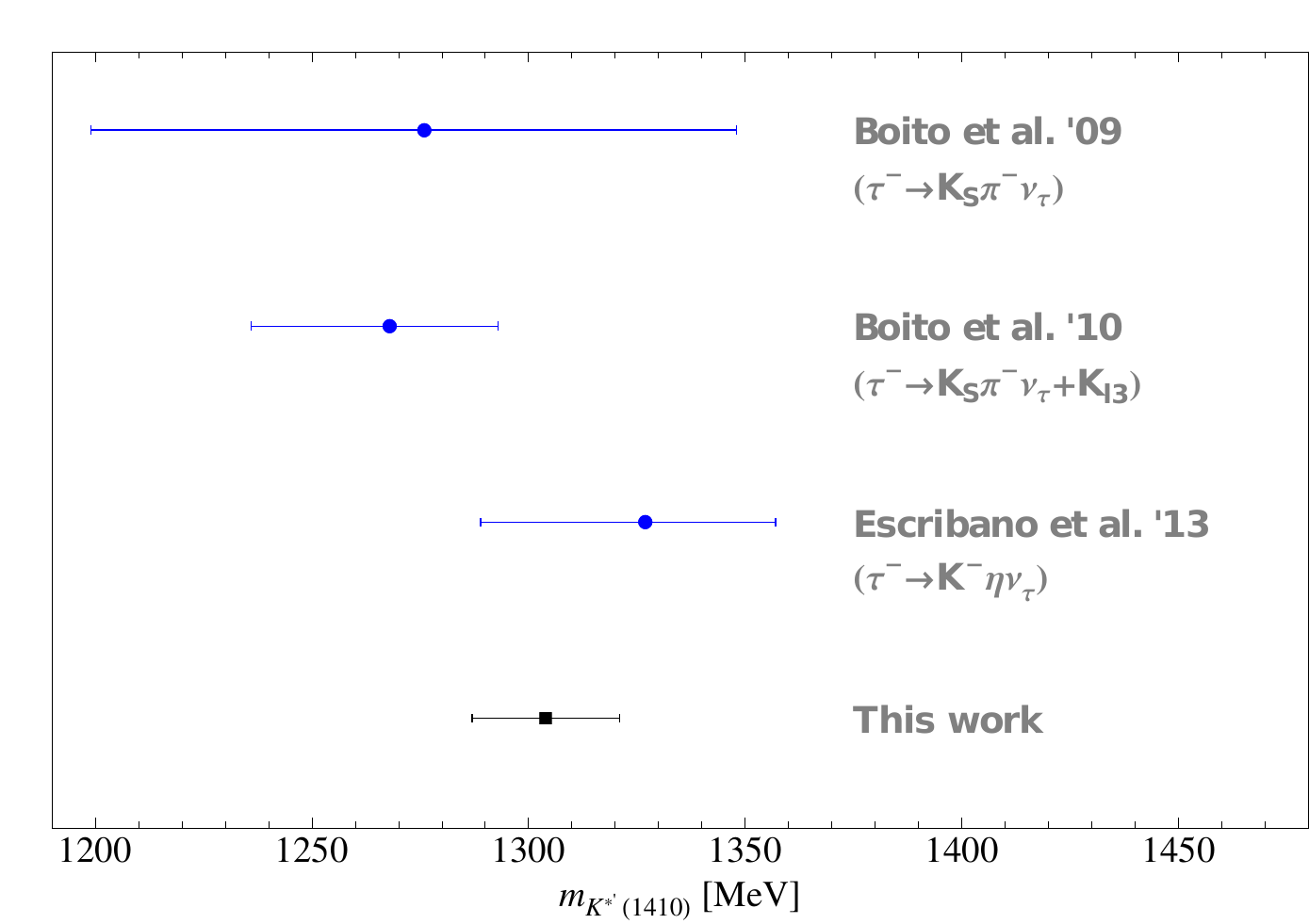}\includegraphics[scale=0.585]{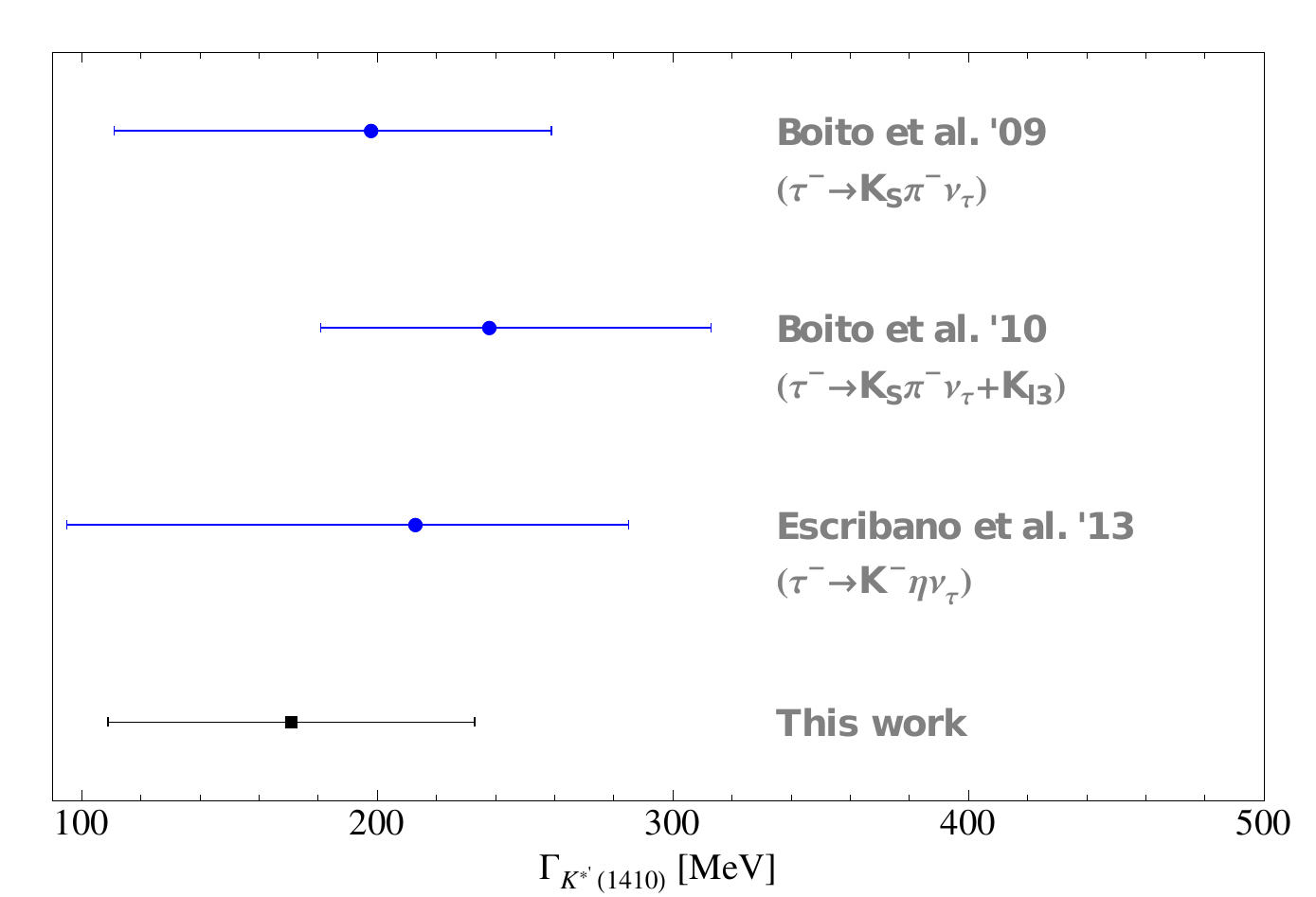}
\caption{\label{values} \small{Our value for the pole parameters, mass (left) and width (right), of the $K^{*}(1410)$ resonance obtained from a joint to both experimental Belle $\tau^-\to K_S\pi^-\nu_\tau$ and  $\tau^-\to K^-\eta\nu_\tau$ decays spectra compared with previous determinations from both channels separated.}}
\end{figure}

Our fits have been performed by relating the experimental Belle $\tau^{-}\to K_{S}\pi^{-}\nu_{\tau}$ and $\tau^{-}\to K^{-}\eta\nu_{\tau}$ spectra with Eq. (\ref{spectral function}) from theory through 

\begin{equation}
\label{theory_to_experiment}
\frac{{\rm d}N_{\rm events}}{{\rm d}\sqrt{s}} \,=\,
\frac{{\rm d}\Gamma(\tau^-\to (PQ)^-\nu_\tau)}{{\rm d}\sqrt{s}}\,
\frac{N_{\rm events}}{\Gamma_\tau \bar{B}(\tau^-\to (PQ)^-\nu_\tau)}
\,\Delta \sqrt{s_{\rm bin}} \,,
\end{equation}
where $N_{\rm events}$ is the total number of events measured for the
considered process, $\Gamma_\tau$ is the inverse $\tau$ lifetime and
$\Delta \sqrt{s_{\rm bin}}$ is the bin width.
$\bar{B}(\tau^-\to (PQ)^-\nu_\tau)\equiv \bar{B}_{PQ}$ is a normalisation
constant that, for a perfect description of the spectrum, would equal the
corresponding experimental branching fraction. Details on these numbers are accurately discussed in Ref. \cite{Escribano:2014joa}. We have to comment here that unfolding of detector effects has not yet been performed for the $\tau^{-}\to K^{-}\eta\nu_{\tau}$ decay. In order to compare the data of both channels on the same footing, we generate a 'pseudounfolding' function from the $K_{S}\pi^{-}$ mode, for which we have both folded and unfolded data, to generate a simulated 'unfolded' data for $\tau^{-}\to K^{-}\eta\nu_{\tau}$. This has been, of course, an assumption, and in order to avoid this approximation in future work it would be really interesting to have unfolded/physical data from the experimental collaborations but, of course, we are willing to provide our codes to the experimental collaborations upon request.
\\
Our central joint fit results are displayed in Table.\ref{tab:a}, where the pole positions have been determined through the standard convention $s_{p}=\left(M_{\rm{pole}}-i\frac{\Gamma_{\rm{pole}}}{2}\right)^{2}$ \cite{Escribano:2002iv}.
\begin{table}
\begin{tabular}{lrrrr}
\hline
  \tablehead{1}{r}{b}{Parameters}& \tablehead{1}{r}{b}{Central values}\\
\hline
$\bar{B}_{K\pi}(\%)$ &  $0.404\pm0.012$ \\
$M_{K^*}$ & $892.03\pm0.19$ \\
$\Gamma_{K^*}$ &  $46.18\pm0.44$ \\
$M_{K^{*\prime}}$ & $1304\pm17$ \\
$\Gamma_{K^*\prime}$ &  $171\pm62$ \\
$\gamma_{K\pi}$ & $=\gamma_{K\eta}$\\
$\lambda^{\prime}_{K\pi}\times10^3$ &  $23.3\pm0.9$\\
$\lambda^{\prime\prime}_{K\pi}\times10^4$ & $11.8\pm0.2$ \\
$\bar{B}_{K\eta}\times10^4$ &  $1.58\pm0.10$\\
$\gamma_{K\eta}\times10^2$ & $-3.4^{+1.2}_{-1.4}$ \\
$\lambda^{\prime}_{K\eta}\times10^3$ & $20.9\pm2.7$ \\
$\lambda^{\prime\prime}_{K\eta}\times10^4$ & $11.1\pm0.5$\\
\hline
\end{tabular}
\caption{Results corresponding to Eq. (3.3) and Eq. (4.1) of Ref. \cite{Escribano:2014joa}. Parameters with dimensions are given in MeV.}
\label{tab:a}
\end{table}
These results correspond to $s_{cut}=4$ GeV$^{2}$ (though the uncertainty associated to its largest variation has been added in quadrature to the statistical fit error, see Ref. \cite{Escribano:2014joa} for further details.) and $\gamma_{K\pi}=\gamma_{K\eta}$. A comparison of our main fit results with the measured Belle distributions is given in Fig. \ref{fig:Spectrum}. 
Comments on our results are in order: we show a nice agreement with the experimental data (corroborated by the $\chi^{2}/n.d.f\sim 1.03$ we have obtained), the $K\pi$ mode is visibly dominated by the $K^{*}(892)$ resonance and both decays are vastly dominated by the vector form factor contribution. Moreover, on one hand the pole parameters of the $K^{*}(892)$ resonance are basically the same than those obtained in Refs. \cite{Boito:2008fq,Boito:2010me} when studying only the $K\pi$ mode. This is what one would have expected since these parameters are driven by the data of the $\tau^-\to K_S\pi^-\nu_\tau$ decay. On the other hand, adding the $\tau^-\to K^-\eta\nu_\tau$ decay mode into the fit we have sizably improved previous determinations of the mass of the $K^{*}(1410)$ resonance while only a slight improvement in the width has been gained. This can be seen in Fig.\ref{values}, where the main result of this work is shown and compared with previous findings.
Regarding the slope parameters, while we have found that the $K_{S}\pi^{-}$ ones are in great accordance with previous analogous determinations \cite{Boito:2008fq,Boito:2010me}, the $K^{-}\eta$ ones show a $2\sigma$ deviation with respect to the $K_{S}\pi^{-}$ ones. This may indicate a possible isospin violation since is the $K^{-}\pi^{0}$ system which enters in describing the $K^{-}\eta$, but it could also be just a statistical effect. In order to disentangle this fact we encourage the experimental collaborations to further investigate the $\tau^-\to K^{-}\pi^{0}\nu_\tau$ channel and publish the corresponding unfolded spectrum. As the required form factor for describing the $K\to\pi l\nu_{l}$ decay is related by crossing symmetry to the $K\pi$ vector form factor of Eq. (\ref{dispersive VFF}), $K_{l3}$ data would also help on this point.

\newpage
\section{Conclusions}
In this work we have exploited our previous experiences on the $\tau^-\to K_S\pi^-\nu_\tau$ and $\tau^-\to K^-\eta\nu_\tau$ decays in separate analyses to perform a joint fit of both channels with the main goal of improving the determination of the pole parameters of the $K^{*}(1410)$. Our central result is a pole mass and width of $M_{K^{*}(1410)}=1304\pm17$ MeV and $\Gamma_{K^{*}(1410)}=171\pm62$ MeV, respectively. We have employed a three-times subtracted dispersive representation for describing the elastic vector form factor which largely dominates both decays. We have found a distinction between the slope parameters depending on the $K\pi$ channel one is looking at, which may indicate a possible isospin violation. We encourage the experimental groups to measure the $\tau^-\to K^{-}\pi^{0}\nu_\tau$ decay to unveil this fact. From the theory side, a coupled channel description for the vector form factor should be done at some point to further improve this study. 

%%%%%%%%%%%%%%%%%%%%%%%%%%%%%%%%%%%%%%%%%%%%%%%%
%% BACKMATTER
%%%%%%%%%%%%%%%%%%%%%%%%%%%%%%%%%%%%%%%%%%%%%%%%

\begin{theacknowledgments}
I want to thank Rafel Escribano, Matthias Jamin, Pere Masjuan and Pablo Roig for helpful comments on the manuscript. This work has been supported in part by the FPI scholarship BES-2012-055371 (S.G-S), the Ministerio de Ciencia e Innovaci\'{o}n under grant FPA2011-25948, the Secretaria d'Universitats i Recerca del Departament d'Economia i Coneixement de la Generalitat de Catalunya under grant 2014
SGR 1450, the Ministerio de Econom\'{i}a y Competitividad under grant SEV-2012-0234, the
Spanish Consolider-Ingenio 2010 Programme CPAN (CSD2007-00042), and the European
Commission under programme FP7-INFRASTRUCTURES-2011-1 (Grant Agreement N.
283286).

\end{theacknowledgments}



\bibliographystyle{aipproc}   % if natbib is available

\bibliography{sample}

\IfFileExists{\jobname.bbl}{}
 {\typeout{}
  \typeout{******************************************}
  \typeout{** Please run "bibtex \jobname" to optain}
  \typeout{** the bibliography and then re-run LaTeX}
  \typeout{** twice to fix the references!}
  \typeout{******************************************}
  \typeout{}
 }

%%%%%%%%%%%%%%%%%%%%%%%%%%%%%%%%%%%%%%%%%%%
%% The following lines show an example how to produce a bibliography
%% without the help of the BibTeX program. This could be used instead
%% of the above.
%%%%%%%%%%%%%%%%%%%%%%%%%%%%%%%%%%%%%%%%%%%

\end{document}